\documentclass[10pt,a4paper]{article}

\usepackage[left=2.5cm, right=2.5cm, top=2.5cm, bottom=2.5cm]{geometry}

\usepackage{hyperref}
\usepackage{settings}
\usepackage{code}
\usepackage[title]{appendix}

\usepackage{array}
\usepackage{pgfplots}
\usepackage{makecell}
\usepackage{relsize}
\usepackage{subcaption}
\usepackage{algpseudocode}

\definecolor{Red}{RGB}{230, 25, 75}
\definecolor{Green}{RGB}{60, 180, 75}
\definecolor{Yellow}{RGB}{255, 225, 25}
\definecolor{Blue}{RGB}{0, 130, 200}
\definecolor{Orange}{RGB}{245, 130, 48}
\definecolor{Purple}{RGB}{145, 30, 180}
\definecolor{Cyan}{RGB}{70, 240, 240}
\definecolor{Magenta}{RGB}{240, 50, 230}
\definecolor{Lime}{RGB}{210, 245, 60}
\definecolor{Pink}{RGB}{250, 190, 190}
\definecolor{Teal}{RGB}{0, 128, 128}
\definecolor{Lavender}{RGB}{230, 190, 255}
\definecolor{Brown}{RGB}{170, 110, 40}
\definecolor{Beige}{RGB}{255, 250, 200}
\definecolor{Maroon}{RGB}{128, 0, 0}
\definecolor{Mint}{RGB}{170, 255, 195}
\definecolor{Olive}{RGB}{128, 128, 0}
\definecolor{Coral}{RGB}{255, 215, 180}
\definecolor{Navy}{RGB}{0, 0, 128}

\usepackage{tikz}
\usetikzlibrary{arrows, fit, positioning}

\definecolor{leafColor}{RGB}{224,255,255} 
\definecolor{innerColor}{RGB}{255,250,205} 
\definecolor{rootColor}{RGB}{255,228,225} 
\definecolor{constantColor}{RGB}{255,222,173} 
\definecolor{lhsInnerColor}{RGB}{240,230,140} 
\definecolor{lhsLeafColor}{RGB}{230,230,250} 
\definecolor{wrapperColor}{RGB}{189,183,107} 

\tikzset{
  nodeBase/.style={circle,draw=black,thick, inner sep=0pt, minimum size=2em, text centered},
  leafNode/.style={nodeBase, fill=leafColor},
  lhsLeafNode/.style={nodeBase, fill=lhsLeafColor},
  innerNode/.style={nodeBase, fill=innerColor},
  lhsInnerNode/.style={nodeBase, fill=lhsInnerColor},
  rootNode/.style={nodeBase, fill=rootColor},
  constantNode/.style={nodeBase, fill=constantColor},
  wrapperNode/.style={nodeBase, fill=wrapperColor},
  mynode/.style={circle,draw=black, fill=leafColor,thick, inner sep=0pt, minimum size=2em, text centered},
  linkArrow/.style={->, >=latex', thick},
  descArrow/.style={->, >=latex', dotted, red, thick}
}

\pgfplotsset{ %
  cycle list={ %
    Red, mark=*\\%
    Blue, mark=square*\\%
    Orange, mark=triangle*\\%
    Green, mark=diamond*\\%
    Maroon, mark=asterisk\\%
    Purple, mark=pentagon*\\%
    Teal, mark=o\\%
    Brown, mark=10-pointed star\\%
    Magenta, mark=Mercedes star\\%
  },
}

\begin{document}

\title{Algorithmic differentiation for domain specific languages in C++ with expression templates}
\author{Max Sagebaum, Nicolas R. Gauger}
\date{Chair for Scientific Computing \\ RPTU University Kaiserslautern-Landau}

\maketitle

\begin{abstract}
The application of operator overloading algorithmic differentiation (AD) to computer programs in order to compute the derivative is quite common. But, the replacement of the underlying computational floating point type with the specialized type of an AD tool has two problems. First, the memory structure of the program is changed and floating-point data is interleaved with identifiers from AD. This prevents the compiler from performing optimizations such as SIMD optimizations. Second, the AD tool does not see any domain-specific operations, e.\,g.\ linear algebra operations, that the program uses. This prevents the AD tool from using specialized algorithms in such places. We propose a new AD tool that is tailored to such situations. The memory structure of the primal data is retained by associating an identifier with each entity, e.\,g.\ matrix, and not with each floating point value, e.\,g.\ element of the matrix. Operations on such entities can then be annotated and a generator is used to create the AD overloads. We demonstrate that this approach provides performance comparable to that of other specializations. In addition, the run-time factor is below the theoretical 4.5 of reverse AD for programs that are written purely with linear algebra entities and operations.
\end{abstract}

\section{Introduction}

Algorithmic differentiation (AD) defines how a function like $F: \R^n \rightarrow \R^m$ can be differentiated in an automatic fashion by applying the chain rule recursively. $F$ is assumed to be a concatenation of elemental functions $\phi_i: \R^k \rightarrow \R$. $i$ ranges from $1$ to $N$, where $N$ can be quite large, e.\,g.\ $1e^9$. Elemental functions arise naturally when an algorithm is implemented as a software and run on a CPU, since in this case only elemental operations like $+$, $*$ or functions like $\sin$ and $\cos$ can be evaluated. The reverse mode of AD for $y = F(x)$ computes
\begin{equation}
  \bar x = \frac{d F}{d x}^\top(x) \bar y \eqdot
  \label{eq:ad_reverse}
\end{equation}
The bar variables $\bar x \in \R^n$ and $\bar y \in \R^m$ are the so called adjoint values. $\bar y$ defines the seeding for the derivative computation. $\bar x$ defines how much the input values need to be changed in order to get a corresponding change of $\bar y$ in the output values $y$. The adjoint $\bar x$ is not evaluated by setting up $\frac{d F}{d x}$. Equation \eqref{eq:ad_reverse} is computed by applying the so-called reverse evaluation process (back propagation). The elemental form of Equation \eqref{eq:ad_reverse} is defined as
\begin{equation}
  \bar w \aeq \frac{d \phi}{d u}^\top(u) \bar w
  \label{eq:ad_elemental_reverse}
\end{equation}
and is applied to every $w_i = \phi_i(u_i)$ in reverse order. That is, from $N$ to $1$.

There are usually two options to apply the reverse mode of AD. Source transformation tools such as TAPENADE \cite{Hascoet2013TTA} or Enzyme \cite{enzyme2020} take the source code of the application and produce new code to compute the adjoint values. The other option consists of operator overloading tools such as dco/c++ \cite{AIB-2016-08}, ADOL-C \cite{Walther2012Gsw}, or CoDiPack \cite{SaAlGauTOMS2019}. They provide an active type like \ic{adouble} that is used in the program instead of the original floating point type. This type overloads all mathematical operators and functions. During the evaluation of the program, it stores some information for each evaluated elemental function on a so-called tape (stack). In the reverse evaluation, this information is read from the tape in reverse order, which allows the computation of Equation \eqref{eq:ad_elemental_reverse} for each elemental function.

The information stored on the tape can rapidly grow for large applications. A growth indicator is the number of elemental functions per second. A matrix vector multiplication usually evaluates $k^2$ multiplications and $k^2 - k$ additions, $k \in \N$ is the size of the matrix. For $k = 10$, a CoDiPack primal value tape would record $2520$ bytes of data. In a specialized setting where matrix and vector entities are natural objects for AD and matrix-vector multiplications are elemental operations, then only $96$ bytes would be required. For the solution of a shifted vector (\ic{M.solve(v2 - v1) + v1}), CoDiPack records $60216$ bytes of data, and a specialized AD tool would record $104$ bytes. These examples indicate that a specialized handling of AD for well-known entities and operations reduces the number of elemental functions and, therefore, the required memory for AD.

Such optimizations have been the target of AD tool implementations for quite some time. The Stan Math Library \cite{carpenter2015stan} introduced several special functions that are optimized for AD reversals. These optimized functions have been partly added to the underlying Eigen framework via template specializations. Also, the 2.0 version of Adept \cite{Hogan2014FRM} features a custom linear algebra implementation where operations have been optimized for AD. Eigen-AD \cite{eigenAD2020} pursues an idea similar to that of Stan Math and Adept. It is a fork of Eigen and the code has been specialized in some places for AD tools. Eigen-AD features a specialized interface that allows arbitrary AD tools to benefit from these specializations. All of these implementations have in common that the active type associates an identifier (AD tool management data) with each floating-point value. A matrix in e.\,g.\ Adept or Eigen-AD contains $n^2$ floating point values and $n^2$ identifiers, which are stored interleaved in memory. A draft paper by the authors \cite{sagebaum2018algorithmic} from 2018 demonstrates an ansatz that wraps e.\,g.\ an entire Eigen matrix in an active type. This reduces the AD management data to one identifier while still storing the $n^2$ floating point entries. In addition, the floating-point values are packed continuously again. The Stan Math Library has implemented a similar approach in 2021. They specialize their AD type for Eigen matrices. Unfortunately, there have been no publications on this improvement and no documentation is available.

In other languages than C++, the same ideas have also been applied. Specialized functions have been implemented in Matlab via ADIMAT \cite{Bischof2002CST}. In Python, AD is driven by machine learning applications. TensorFlow \cite{tensorflow2015-whitepaper}, PyTorch \cite{Ansel2024}, or JAX \cite{jax2018github} are prominent examples. These tools use the language reflection mechanism to compile just-in-time improved versions of the AD code. There are also several other examples in other languages.

A domain specific language (DSL) is a set of entities like matrices or vectors, together with operations on these entities like multiplications or linear system solves. Complex numbers or SIMD vectors are examples for common DSLs. Neural networks for machine learning can be seen as DSLs as well. As first demonstrated in \cite{sagebaum2018algorithmic}, we want to create an AD tool for arbitrary DSLs in C++ using expression templates \cite{Hogan2014FRM, SaAlGauTOMS2019}. One design focus for the tool is the simple integration of new DSLs by users.

In this paper, we will focus on the design of the AD tool itself and only briefly highlight the expression generation. Section \ref{sec:ad_expressions} introduces the extensions to AD expression template implementations for handling arbitrary DSLs. Afterwards, the design of the AD tool is discussed in Section \ref{sec:ad_tool_design}. The source code annotations required on a DSL for the AD tool are introduced in Section \ref{sec:ad_dsl_definition}, and examples for the Eigen library \cite{eigenweb} are given. Finally, the performance of the AD tool is evaluated on synthetic benchmarks.

\section{AD Expressions}
\label{sec:ad_expressions}

The reverse AD equation for the elemental operations defined in Equation \eqref{eq:ad_elemental_reverse} shows how the adjoint values are propagated. $\phi$ is usually defined with only one output since this is true for all real-valued elemental operations. Griewank and Walther extend this definition in \cite[page 25f]{grie08} to an arbitrary number of outputs. Therefore, we define $\phi_i$ as $\phi_i: \R^k \rightarrow \R^l$, where $l$ describes the dimension of the outputs.

The state-of-the-art generation of elemental functions for operator overloading AD tools is currently done by expression templates \cite{Hogan2014FRM, SaAlGauTOMS2019}. An assignment like $w = (a + b) * (c - d)$ generates an expression template like:
\begin{code}
MulExpr<AddExpr<AdType, AdType>, SubExpr<AdType, AdType>>
\end{code}
\vspace{-0.75em}
This template encodes all the information about the assignment. The AD tool can then infer information from the expression template such as derivative values. Figure \ref{fig:exp_basic} shows a graph representation of the expression template, which is simpler to understand and compare. It consists of leafs, nodes, and links. The graph always has a root node that represents the single output of the expression.

For real-valued statements, only right-hand side (rhs) expressions are required. In the context of DSLs, left-hand side (lhs) expressions also need to be added to the expression framework. We will use the expression
\begin{equation}
  v[1] = a * b[0] / 2.0
  \label{eq:example_expression}
\end{equation}
as an example throughout the discussion. The operator $[\cdot]$ represents the array access to a vector.

The graph of equation \eqref{eq:example_expression} is shown in Figure \ref{fig:example_expression_graph}, on which we will discuss three properties. Firstly, we now also have to consider the assignment as part of the expression as well as the branch for access to the lhs value, which is the most general representation. Having separate expressions for the lhs and rhs part would seem simpler, but this would exclude operations such as the update operator$\aeq$ or functions like \ic{func(A const& a_in, W& w_out, V& v_out)} with multiple output values. For this reason, we assume that each expression has only one root node and that this root node has at least one lhs branch and one rhs branch. We call such an expression a \emph{compute} expression. Secondly, we also assume that the compute expressions are not nested. That is, each compute expression is stored on the tape before the return value is created. An example would be \ic{z *= v *= w *= a}, which includes three compute expressions. (In C++, all assignment operators return the lhs value.) Such an expression could be handled as one, but the implementation is quite involved. Instead, we store three separate expressions on the tape.

Lastly, constant values are also included in the expression. The array access expression has therefore two child expressions, one for the vector and one for the index into the vector. This setup simplifies the handling in the AD tool. It only has to consider leaf values and does not need to check for each node if additional values like the access index have to be stored. This choice also simplifies the implementation of expression templates.
\begin{figure}
  \begin{minipage}[b]{0.49\textwidth}
    \begin{center}
      \begin{tikzpicture}[node distance=1cm]

        \node[rootNode] (mul) {$*$};
        \node[above=1.5cm of mul.center] (l1) {};
        \node[innerNode, left=0.8cm of l1.center] (add) {$+$};
        \node[above=1.5cm of l1.center] (l2) {};
        \node[leafNode, left=1.45cm of l2.center] (a) {$a$};
        \node[leafNode, left=0.15cm of l2.center] (b) {$b$};
        \node[innerNode, right=0.8cm of l1.center] (sub) {$-$};
        \node[leafNode, right=0.15cm of l2.center] (c) {$c$};
        \node[leafNode, right=1.45cm of l2.center] (d) {$d$};
        \draw[linkArrow] (mul) -- (add) node [midway] (link1) {};
        \draw[linkArrow] (mul) -- (sub);
        \draw[linkArrow] (add) -- (a) node [midway] (link2) {};
        \draw[linkArrow] (add) -- (b);
        \draw[linkArrow] (sub) -- (c);
        \draw[linkArrow] (sub) -- (d);

        \node[below right=0.5cm and -0.75cm of sub] (node) {Rhs node};
        \node[left=0.5cm of link1] (link) {Link};
        \node[left=0.5cm of mul] (root) {Root};
        \node[below left=0.5cm and -0.5cm of c] (leaf) {Rhs leaf};

        \draw[descArrow] (root) -- (mul);
        \draw[descArrow] (node) -- (sub);

        \draw[descArrow] (link) -- (link1);

        \draw[descArrow] (leaf) -- (c);
      \end{tikzpicture}
    \end{center}

    \caption{Graph representation of the generated expression template for $w = (a + b) * (c - d)$}
    \label{fig:exp_basic}
  \end{minipage}
  \hfill
  \begin{minipage}[b]{0.49\textwidth}
    \begin{center}
      \begin{tikzpicture}[node distance=1cm]
        \node[rootNode] (assign) {$=$};
        \node[above=0.5cm of assign.center] (l1) {};

        \node[lhsInnerNode, left=0.8cm of l1.center] (array_v) {$[\cdot]$};
        \node[above=0.6cm of array_v.center] (l2_left) {};
        \node[lhsLeafNode, left=0.3cm of l2_left.center] (v) {$v$};
        \node[constantNode, right=0.3cm of l2_left.center] (index_v) {$1$};

        \node[innerNode, right=0.8cm of l1.center] (div) {$/$};
        \node[above=0.6cm of div.center] (lt_right) {};
        \node[constantNode, right=0.3cm of lt_right.center] (div_2) {$2.0$};

        \node[innerNode, left=0.4cm of lt_right.center] (mul) {$*$};
        \node[above=0.7cm of mul.center] (l2_right) {};
        \node[leafNode, left=0.3cm of l2_right.center] (a) {$a$};
        \node[innerNode, right=0.3cm of l2_right.center] (array_b) {$[\cdot]$};
        \node[above=0.6cm of array_b.center] (l3) {};

        \node[leafNode, left=0.4cm of l3.center] (b) {$b$};
        \node[constantNode, right=0.4cm of l3.center] (index_b) {$0$};

        \node[left=0.5cm of array_v] (lhs_node) {Lhs node};
        \node[above=0.5cm of v] (lhs_leaf) {Lhs leaf};
        \node[below=0.75cm of div_2] (constant) {Constant};

        \draw[linkArrow] (assign) -- (array_v);
        \draw[linkArrow] (assign) -- (div);

        \draw[linkArrow] (array_v) -- (v);
        \draw[linkArrow] (array_v) -- (index_v);

        \draw[linkArrow] (div) -- (mul);
        \draw[linkArrow] (div) -- (div_2);

        \draw[linkArrow] (mul) -- (a);
        \draw[linkArrow] (mul) -- (array_b);

        \draw[linkArrow] (array_b) -- (b);
        \draw[linkArrow] (array_b) -- (index_b);

        \draw[descArrow] (lhs_node) -- (array_v);
        \draw[descArrow] (lhs_leaf) -- (v);
        \draw[descArrow] (constant) -- (div_2);
      \end{tikzpicture}
    \end{center}
    \caption{Graph representation of the generated expression template for Equation \ref{eq:example_expression}.}
    \label{fig:example_expression_graph}
  \end{minipage}
\end{figure}

In this paper, we only briefly present three highlights for the generation of expression templates for DSLs. A more complete presentation will be provided in a follow-up paper.

The generation of expression templates for e.\,g.\ Eigen or complex numbers makes it necessary to add member operations to the expression. For example, if $a$ and $b$ are complex values, then \ic{(a * b).real()} is a valid statement. In the context of expressions, the same statement should also work, meaning that member functions need to be injected into the expressions. This is achieved by adding the interface \ic{MemberOperations<Value, Impl>} to the inheritance hierarchy of the expression templates. The class \ic{MemberOperations} can be specialized for the template argument \ic{Value}. This specialization implements the member functions, which will make them accessible in all expressions that return the value. \ic{Impl} refers to the child class at the top of the inheritance chain and comes from the curiously recurring template pattern used by expression templates.

Second, there can be multiple \ic{MemberOperations} in the inheritance hierarchy. If \ic{B} inherits from \ic{A}, then \ic{MemberOperations<B, Impl>} can inherit from \ic{MemberOperations<A, Impl>}. This will make all member functions of \ic{A} and \ic{B} available in expressions of type \ic{B}.

Third, the adjoint computation in the expression needs to be considered. For every input argument of an operation, e.\,g.\ "val", a function \ic{diff_val(Adj val_b, R_b r_b, <primal values>)} is generated. \ic{r_b} is the adjoint value from the return value of the expression and can be used to compute the adjoint/derivative with respect to the argument. Instead of returning the computed adjoint, it can be applied via the object \ic{val_b}, e.\,g.\ \ic{val_b += 2.0 * val * r_b}. This has the advantage that other values used for the computation of the adjoint are not deconstructed on the return, which avoids dangling pointers. The \ic{+=} operation will open a new function context, and all intermediate variables are still accessible. This is important for Eigen, since Eigen captures its expressions by reference. They would contain dangling pointers if the intermediate values are destroyed on return.

\section{DSL AD tool design}
\label{sec:ad_tool_design}

Several design choices must be made to implement the DSL AD tool. In this section, we want to explain the reasoning behind most of the decisions. The current implementation is nearly self-contained. Only some basic infrastructure from CoDiPack is used. There are two reasons for this decision. First, CoDiPack is fixed at the \icNC{C++11} standard. The new tool requires features from the \icNC{C++20} standard. Second, the expressions in the new tool require more features than those from CoDiPack. Therefore, a new implementation, without the legacy interface from CoDiPack, simplifies the exploration of new ideas for the extended-expression templates. In the long run, the DSL AD tool will be integrated into CoDiPack.

For operator overloading AD tools there are two main implementation strategies, Jacobian taping \cite{SaAlGauTOMS2019} and primal value taping \cite{SaAlGa2018OMS}. Jacobian taping is very simple to implement since only the Jacobian $\frac{d \phi}{d u}$ of $\phi$ must be stored. In the real value case $l$ is always one, making the Jacobian a vector of size $k$. Memory consumption is therefore linear. In the DSL case $l$ can be greater than one and the curse of dimensionality applies. For example, for a vector-valued function, the memory consumption for storing the Jacobian is quadratic, and for a matrix-valued function the memory consumption is cubic. Therefore, we chose a primal value taping approach for our tool where the lhs values are stored on the tape. This also has the advantage that no extra entities, like the Jacobian of a matrix with respect to a matrix, are stored on the tape as would be the case in the Jacobian setting.

There are two options to manage the identifiers. A linear one, described in \cite[Section 6.1]{grie08} or employed by dco/c++ \cite{AIB-2016-08}, or a reuse one, described in \cite[Section 4.1]{grie08} or used by ADOL-C \cite{Walther2012Gsw}. In terms of memory to store the primal values, both schemes are nearly identical. However, the size of the adjoint vector between both varies. In the linear case, all adjoint entities are kept during the reverse evaluation. A reuse scheme overwrites the adjoint values, making the size of the adjoint vector usually smaller, as demonstrated in \cite{sagebaum2023assign}. Since we need to store entities of higher order in the adjoint vector, a reuse index management is chosen as the default scheme.

The definition of the AD-type implementation for operator overloading is shown in Figure \ref{fig:ad_type_definition}. In contrast to Adept \cite{Hogan2014FRM} or the default implementation in the Stan Math Library \cite{carpenter2015stan}, we do not have one identifier for each real-valued entry but one identifier for the entire entity. This has two advantages. Firstly, the primal value entity \ic{Value} is not changed. We do not introduce the identifier into its structure. Optimizations such as SIMD access are still possible. Second, only one identifier needs to be stored per argument, and not one for each element. One disadvantage is the lack of access to raw data. If each real value had an identifier, access to the raw data would directly yield the proper AD types. However, this disadvantage can usually be overcome by specialized expression implementations.

Primal value tape implementations usually have one vector for all primal values and a vector for all adjoint values. Since we choose to identify a whole entity with one identifier, it makes sense to have dedicated primal and adjoint vectors for each entity. Therefore, we also need a dedicated identifier manager for each kind of entity. This means that there will be one for double values, vector values, and matrix values. However, it is not necessary to connect the identifiers to the corresponding type. Each expression knows the type of each argument and will therefore know the corresponding type of each identifier. Figure \ref{fig:tape_memory_layout} visualizes the proposed memory layout for the vectors. If an application uses e.g. Eigen matrices, then \ic{Eigen::Matrix<2,2>} and \ic{Eigen::Matrix<Dynamic, Dynamic>} are registered as two different types. This simplifies the management of dynamic types, like the dynamic Eigen matrix, since they are not mixed with fixed-size types. A set to zero on a dynamic type has to reset the data structure such that this can be recognized. The first update operation ($\aeq$) on an adjoint value has to resize the dynamic type to the size of the update. All subsequent updates should have the same size.

\begin{figure}
  \begin{minipage}[b]{0.49\textwidth}
    \begin{center}
      \begin{tikzpicture}[node distance=1cm]
          \tikzset{
            fitting node/.style={
              inner sep=0pt,
              fill=none,
              draw=none,
              reset transform,
              fit={(\pgf@pathminx,\pgf@pathminy) (\pgf@pathmaxx,\pgf@pathmaxy)}
            },
            reset transform/.code={\pgftransformreset}
          }
          \tikzset{
              mymark/.style={},
              myarrow/.style={->, >=latex', shorten >=1pt, thick},
              myarrow2/.style={->, >=latex', shorten >=1pt, thick, dashed},
              mylabel/.style={text width=7em, text centered}
          }
          \newcommand\IndexColor{blue}

          \node[] (global_vectors) {Global vectors (RAM):};

          \node[below=0.6cm of global_vectors.west, anchor=west]
           (vector_data) {Vector:};

          \node[below=0.5cm of vector_data.west, anchor=west]
           (primal) {Primal};
          \node[right=0.0cm of primal.east, anchor=west]
                      (adjoint) {Adjoint};

          \draw[black] (primal.south) ++(-0.25cm, -0.25cm) rectangle ++(0.5cm, -2cm);
          \draw[black] (adjoint.south) ++(-0.25cm, -0.25cm) rectangle ++(0.5cm, -2cm);
          \node[below=3.5cm of vector_data.west, anchor=west] (index_manager) {Index manager};
          \draw[black] (index_manager.south) ++(-0.2cm, 0.0cm) rectangle ++(1.3em, -1.3em) node (temp) {};

          \node[draw,black, fit=(vector_data) (index_manager) (temp)] (vector_data_fit) {};

          \node[right=1.5cm of vector_data.east, anchor=west]
           (matrix_data) {Matrix:};

          \node[below=0.5cm of matrix_data.west, anchor=west]
           (primal) {Primal};
          \node[right=0.0cm of primal.east, anchor=west]
                      (adjoint) {Adjoint};

          \draw[black] (primal.south) ++(-0.25cm, -0.25cm) rectangle ++(0.5cm, -2cm);
          \draw[black] (adjoint.south) ++(-0.25cm, -0.25cm) rectangle ++(0.5cm, -2cm);
          \node[below=3.5cm of matrix_data.west, anchor=west] (index_manager) {Index manager};
          \draw[black] (index_manager.south) ++(-0.2cm, 0.0cm) rectangle ++(1.3em, -1.3em) node (temp) {};

          \node[draw,black, fit=(matrix_data) (index_manager) (temp)] (matrix_data_fit) {};

          \node[draw,black, right=0.05cm of matrix_data_fit, minimum height=13.7em] (other_data_fit) {$\ldots$};

          \node[draw,black, fit=(global_vectors) (vector_data_fit) (matrix_data_fit) (other_data_fit)] (global_data_fit) {};

          \node[below=5.1cm of vector_data.west, anchor=west] (streams_data) {Streams (SAM):};

          \node[below=0.25cm of streams_data.south east] (belowStream) {};
          \node[left=-0.55cm of belowStream, anchor=east] (1StreamName) {Expr. handle:};
          \node[below=0.55cm of 1StreamName.east, anchor=east] (2StreamName) {Data size:};
          \node[below=0.55cm of 2StreamName.east, anchor=east] (3StreamName) {Byte data:};

          \draw[black] (1StreamName.north east) ++(0.1cm, -0.05cm) rectangle ++(1.8cm, -1.3em);
          \draw[black] (2StreamName.north east) ++(0.1cm, -0.05cm) rectangle ++(2.2cm, -1.3em);
          \draw[black] (3StreamName.north east) ++(0.1cm, -0.05cm) rectangle ++(2.81cm, -1.3em) node (temp) {};

          \node[left=-0.12cm of streams_data.west] (temp2) {};

          \node[draw,black, fit=(streams_data) (3StreamName) (temp) (temp2)] (global_data_fit) {};
        \end{tikzpicture}
      \end{center}
      \caption{Memory layout of the tape. Global vectors are allocated for each registered DSL entity. Sequential access data is kept general such that arbitrary data can be stored.}
      \label{fig:tape_memory_layout}
  \end{minipage}
  \hfill
  \begin{minipage}[b]{0.49\textwidth}
    \begin{code}
template<typename Value>
struct ActiveType : public <expression base> {
  Value value;
  int identifier;
};
    \end{code}
    \caption{Definition of the overloaded AD type.}
    \label{fig:ad_type_definition}
    \vspace{1em}
    \begin{center}
      \begin{tikzpicture}[node distance=1cm]
        \node[rootNode] (assign) {$=$};
        \node[above=0.5cm of assign.center] (l1) {};

        \node[wrapperNode, left=0.8cm of l1.center] (wrapper) {Wr};
        \node[above=1.5cm of wrapper.center] (lw) {};
        \node[above=0.5cm of wrapper.center] (lw2) {};
        \node[constantNode, left=0.8cm of lw2.center] (old) {Old};

        \node[lhsInnerNode, left=0.4cm of lw.center] (array_v) {$[\cdot]$};
        \node[above=0.6cm of array_v.center] (l2_left) {};
        \node[wrapperNode, left=0.3cm of l2_left.center] (wrap_v) {Id};
        \node[lhsLeafNode, above=0.7cm of wrap_v.center] (v) {$v.id$};
        \node[constantNode, right=0.3cm of l2_left.center] (index_v) {$1$};

        \node[innerNode, right=0.8cm of l1.center] (div) {$/$};
        \node[above=0.6cm of div.center] (lt_right) {};
        \node[constantNode, right=0.3cm of lt_right.center] (div_2) {$2.0$};

        \node[innerNode, left=0.4cm of lt_right.center] (mul) {$*$};
        \node[above=0.7cm of mul.center] (l2_right) {};
        \node[wrapperNode, left=0.5cm of l2_right.center] (wrap_a) {Id};
        \node[leafNode, above=1.45cm of wrap_a.center] (a) {$a.id$};
        \node[innerNode, right=0.3cm of l2_right.center] (array_b) {$[\cdot]$};
        \node[above=0.6cm of array_b.center] (l3) {};

        \node[wrapperNode, left=0.4cm of l3.center] (wrap_b) {Id};
        \node[leafNode, above =0.75cm of wrap_b.center] (b) {$b.id$};
        \node[constantNode, right=0.4cm of l3.center] (index_b) {$0$};

        \draw[linkArrow] (assign) -- (wrapper);
        \draw[linkArrow] (assign) -- (div);

        \draw[linkArrow] (wrapper) -- (array_v);
        \draw[linkArrow] (wrapper) -- (old);

        \draw[linkArrow] (array_v) -- (wrap_v);
        \draw[linkArrow] (wrap_v) -- (v);
        \draw[linkArrow] (array_v) -- (index_v);

        \draw[linkArrow] (div) -- (mul);
        \draw[linkArrow] (div) -- (div_2);

        \draw[linkArrow] (mul) -- (wrap_a);
        \draw[linkArrow] (wrap_a) -- (a);
        \draw[linkArrow] (mul) -- (array_b);

        \draw[linkArrow] (array_b) -- (wrap_b);
        \draw[linkArrow] (wrap_b) -- (b);
        \draw[linkArrow] (array_b) -- (index_b);
      \end{tikzpicture}
    \end{center}
    \caption{Reconstructed expression graph for the example Equation \eqref{eq:example_expression}. A wrapper node for the old value is inserted and the leafs are exchanged with id nodes.}
    \label{fig:expression_reverse_example}
    \vspace{1.2em}
  \end{minipage}
\end{figure}

Figure \ref{fig:tape_memory_layout} also shows the memory layout of the stream data, which is discussed now. The primal value implementation used by ADOL-C \cite{Walther2012Gsw} and the one proposed for CoDiPack \cite{SaAlGa2018OMS} use different data streams for each kind of data required by the tape, e.\,g.\ identifier data, primal values, operators. Since we do not know upfront which entities are required, only three data streams are used. The stream that holds the handles to the expression, the one that stores the data size for each expression, and a byte stream where each expression can store arbitrary data. This design simplifies the data management of the tape. It also does not pose restrictions on the data in the byte stream. The tape loads the data for the expression and then the expression can access the data as it wants. Faulty expression implementations, such as out-of-bounds access, can then be detected by the tape.

Since the data structures are now in place, we can discuss what kind of data entries are stored for each expression. It is mandatory that an identifier is stored for each active argument. For Equation \eqref{eq:example_expression}, these would be the identifiers for $v$, $a$, and $b$. We also need to store the values of constants, which would be $1$, $0$, and $2.0$. In a reuse index management approach, it is also required to restore the old primal value of the lhs. Therefore, for each lhs value we have to store the old primal value of the assigned identifier. These are the basic data entries for each expression. Additional data is required if an online activity analysis \cite{SaAlGauTOMS2019} is performed. Here, the identifier of an active type can be zero. This is called a passive value. Since the zero identifier is shared by all passive values, the current primal value needs to be stored in the byte data. Special care has to be taken for passive values that are lhs values but are also used as rhs values at the same time. An example is \ic{ w *= b} where $w$ is used as a rhs value and a lhs value. $w$ will be give a new identifier, and the old value of this identifier is stored. In addition, the current value of $w$ needs to be stored, since it is required in the evaluation. During reverse evaluation,
first the current value of $w$ needs to be restored and after reverse evaluation of the expression, the old value of the identifier of $w$ needs to be restored.

One memory optimization can be performed for the access of lhs values. The above definition would store the entire vector $v$ for the access of $v[1]$ in Equation \eqref{eq:example_expression}. Since only one element is updated, the other elements would be overwritten with the same values. If only the modified entries are stored, a lot of memory can be saved, especially for larger entities. Therefore, the primal value of each lhs expression is stored at the root of the compute expression, since this is the point where each lhs value is modified. For Equation \eqref{eq:example_expression}, only the old value of $v[1]$ is stored instead of the whole vector $v$. For an lhs expression where this is not true, this behavior can be disabled on a per-node basis. This optimization is incompatible with a copy optimization strategy \cite{AIB-2016-08, sagebaum2023assign}. Here, a new identifier is always created for each lhs, which would require a full store of the old entry in the primal value vector.

This concludes the discussion on the major design choices. The remaining discussion will focus on the reverse evaluation of one compute expression. For each compute expression, the function \ic{store} is called. Here, all data entries are stored as described above. In addition, the size of the data and a function handle are stored. The function handle is created from a static template function \ic{eval_reverse<Expr>} where \ic{Expr} is the type of the compute expression. A call to this function gives us access to the type of the compute expression and therefore the graph depicted in Figure~\ref{fig:example_expression_graph}, but not the original data captured in this expression, e.\,g.\ the values and identifiers. These must be restored from the byte data stream. The reverse evaluation process for one compute expression is as follows:
\begin{enumerate}
  \item Create an expression by:
  \begin{itemize}
    \item reading an identifier for each rhs leaf, reading the primal value if the value is passive and replacing the leaf with a tape active value leaf.
    \item reading an identifier for each lhs leaf and replacing the leaf with a tape active value leaf.
    \item reading the value for each constant leaf.
    \item reading the old primal value for each lhs expression root and reading the current primal value if the value was passive and is also accessed as a rhs value.
  \end{itemize}
  The resulting reverse expression graph for Equation \eqref{eq:example_expression} is shown in Figure \ref{fig:expression_reverse_example}.
  \item Perform the reverse evaluation by:
  \begin{itemize}
    \item getting the adjoint expression of each lhs expression root, store the adjoint value of the expression and reset the adjoint value of the expression to zero.
    \item propagate all adjoint values to each rhs expression root.
  \end{itemize}
  It is important to note that each adjoint is handled at the root of the lhs expression and not at the leaf. Only the selected part of the adjoint value is set to zero, and not the whole value.
  \item Restore the old primal value for each lhs expression root when the value was passive and is also accessed as a rhs value.
\end{enumerate}

\section{Annotations for DSL AD}
\label{sec:ad_dsl_definition}

\begin{table*}
  \begin{tabular}{ll}
  \hline
  Name & Description \\
  \hline
  \ic{AD_IN(name, adj)} & Input parameter: \ic{adj} defines the adjoint computation,\\
   & e.\,g.\ \ic{name_b += name * r_b}. \\
  \ic{AD_OUT(name, acc)} & Output parameter: \ic{acc} defines the adjoint access, e.\,g.\ \ic{name_b[index]}. \\
  \ic{AD_INOUT(name, adj, acc)} & Input and output parameter: \ic{adj} and \ic{acc} are the same as above. \\
  \ic{AD_PASSIVE(name)} & Passive input: No derivative is computed with respect to this argument. The\\
   & primal value is stored. \\
  
  \ic{AD_ELE_PASSIVE} & Passive function: No derivatives are computed for this function. \\
  \ic{AD_INP(name)} & Input parameter for passive function: Primal value is extracted. \\
  \ic{AD_OUTP(name)} & Output parameter for passive function: Primal value is extracted and set to\\
   & passive afterwards. \\
  \ic{AD_INOUTP(name)} & Input and output parameter: Same as above. \\
  \hline
  \end{tabular}
  \caption{Common annotations for defining the DSL AD tool.}
  \label{tab:annotations}
\end{table*}

Using the new DSL AD tool requires a certain workflow, which we want to describe here. For an already finished DSL only the general header \ic{<dslADTool.hpp>} and the DSL specific header, e.\,g.\ \ic{<dslAD_eigen.hpp>} for Eigen, need to be included. The \emph{Hello World} example is shown in Appendix \ref{sec:hello_world}. The general workflow for registering input, output, and setting gradient data is comparable to other AD tools. The major difference occurs in line 9 where the \ic{double} type is registered on the tape. This allows the tape to manage the primal and adjoint vector data for the user, e.\,g.\ resizing the adjoint to the correct size.

When the user wants to integrate a new DSL into the AD tool, the process becomes more involved. First, the code of the DSL must be annotated. Second, the annotated code needs to be parsed by a preprocessor that generates the header file with the expressions. Third, the generated header needs to be included in the project. The first step, annotating the source code, can either be in-place, meaning the original source code of the DSL is annotated, or external, where a separate header is annotated. Depending on the use case, either option has advantages. The most common AD-specific annotations for the source code are listed in Table \ref{tab:annotations}, and an example is given in Figure \ref{fig:annotation_example}. Here, \ic{self} always describes the primary value of the structure. \ic{r_b} represents the adjoint of the return value. After the annotation is complete, the second step needs to be done. The tool \ic{dslADGenerator} is developed separately for this purpose and will be described in the follow-up paper about the expression templates. It reads the annotated source code and generates the header with the expression templates. The generated header can then be used in combination with the DSL AD tool for derivative computations with the annotated DSL. The third step is then the same as above. The generated header is used for the derivative computation.

The current implementation of \ic{dslADGenerator} requires that everything has to be annotated. In the future, more and more defaults will be added. This will make the annotation process less cumbersome and less verbose.

\subsection{DSL AD for Eigen}
We are planning to have a nearly complete annotation of the Eigen library. Currently, the functions for the test cases are annotated, and the most common interface functions. This includes block access, column/row access, some component-wise functions, and some solve functions. An estimate on the coverage can currently not be given, but most of the common use cases should work.

The \ic{ActiveType} design in Figure \ref{fig:ad_type_definition} would require the replacement of all Eigen types with the active type, e.\,g.\ \ic{Eigen::Matrix<Scalar, 3, 3>} with \ic{ActiveType<Eigen::Matrix<Scalar, 3, 3>>}. Even if \ic{Scalar} is already an active double type, the DSL AD specific expressions for Eigen would not be called. The replacement is still necessary. Since Eigen types are template types, a general solution can be implemented through template specialization, which is shown in Figure \ref{fig:eigen_active_type_definition}. This specialization replaces an Eigen matrix of an active type with the active type of an Eigen matrix with the underlying scalar type of the active type. With this specialization, only the scalar type of the matrix needs to be an active type, and no additional replacements are required.
\begin{figure}
  \begin{minipage}[b]{0.49\textwidth}
    \begin{code}
struct Matrix {
  AD_IN(self, self_b += r_b * o.transpose())
  AD_IN(o, o_b += self.transpose() * r_b)
  Matrix operator*(Matrix const& o) const;
  
  AD_INOUT(self, self_b += r_b, self_b(r, c))
  AD_PASSIVE(r) AD_PASSIVE(c)
  double& operator()(int r, int c);
  
  AD_ELE_PASSIVE AD_INP(self)
  int size() const;
}
    \end{code}
    \caption{DSL AD annotation example for the matrix multiplication, the element access and the size. \ic{r_b} is the adjoint value of the return value.}
    \label{fig:annotation_example}
    \vspace{1.2em}
  \end{minipage}
  \hfill
  \begin{minipage}[b]{0.49\textwidth}
    \begin{code}
// namespace Eigen
template<typename Scalar, ...>
struct Matrix<ActiveType<Scalar, int>, ...> :
    public ActiveType<Matrix<Scalar, ...>, int>
{
  using Base = ActiveType<Matrix<Scalar, ...>,
                          int>;
  using Base::Base;
  using Base::operator=;
};
    \end{code}
    \caption{Template specialization for Eigen matrices. A matrix with an active type as a scalar is specialized to be an active type of a matrix. (\ic{...} is a placeholder for the additional template arguments of \ic{Eigen::Matrix}.)}
    \label{fig:eigen_active_type_definition}
  \end{minipage}
\end{figure}

\section{Performance tests}
\label{sec:results}

\subsection{Coupled Burgers' equations}

The coupled Burgers' equations are used to compare the performance of the DSL AD tool with CoDiPack and the Stan Math Library. The setup and discretization of the problem are already described in \cite{SaAlGauTOMS2019} and is also used, e.\,g., in \cite{bluehdorn2023Event}. For completeness, we briefly recapitulate the problem formulation here.

The coupled Burgers' equation \cite{biazar2009exact,bahadir2003fully,zhu2010numerical}
\begin{align}
  u_t + uu_x + vu_y &= \frac{1}{R}(u_{xx} + u_{yy}), \\
  v_t + uv_x + vv_y &= \frac{1}{R}(v_{xx} + v_{yy})
\end{align}
is discretized with an upwind finite difference scheme.
The initial and boundary conditions are taken from the exact solution
\begin{align}
  u(x, y, t) &= \frac{x + y - 2xt}{1 - 2t^2} \quad (x,y,t) \in D \times \R, \\
  v(x, y, t) &= \frac{x - y - 2yt}{1 - 2t^2} \quad (x,y,t) \in D \times \R
\end{align}
given in \cite{biazar2009exact}.
The computational domain $D$ is the unit square $D = [0,1] \times [0,1] \subset \R \times \R$.
As far as the differentiation is concerned, we choose the initial solution of the time-stepping scheme as input parameters, and as the output parameter, we take the squared norm of the final solution.

The timing values are averaged over 20 evaluations. In addition, the CPU frequency is fixed. This stabilizes the time measurements, which are run on one node of the Elwetritsch cluster at the University of Kaiserslautern-Landau (RPTU).
The node consists of two AMD EPYC 7262 CPUs with a total of 16 cores and 256 GB of main memory. We show results only where one process is run on the entire node. Measurements using 16 cores lead to equivalent results.

Figure \ref{fig:burgers_result} shows the results for 32 time steps. The grid size ranges from 500 to 1000. The DSL AD tool performs similar to the CoDiPack primal value type. The recording time is slightly better. In the reverse evaluation, no real difference can be seen between the two tools. The memory graph shows that the DSL AD tool requires slightly more memory than the CoDiPack primal value tape. Thus, the new DSL AD tool has performance similar to that of an equivalent implementation in CoDiPack. Due to the more general nature of the new tool, a slight overhead in memory can be seen. The Jacobian taping approach in CoDiPack usually performs better than primal value tape implementations, which can be seen in the recording and reversal times. The Stan Math Library type does not use expressions templates, which results in higher overall numbers.

\newcommand{\timelinePlot}[4]{
	\addplot+[memoryDefaults] table[x index=#1, y index=#2] {#4};
	\addlegendentry{#3}
}

\newcommand{\timelinePlotWithMark}[4]{
	\addplot+[memoryDefaultsWithMark] table[x index=#1, y index=#2] {#4};
	\addlegendentry{#3}
}

\pgfplotsset{memoryDefaults/.style={
  line width=1.0,
  no marks
}}

\pgfplotsset{memoryDefaultsWithMark/.style={
  line width=1.0,
  mark=o,
  mark size=2pt,
  mark options={
    line width=0.75pt}}}

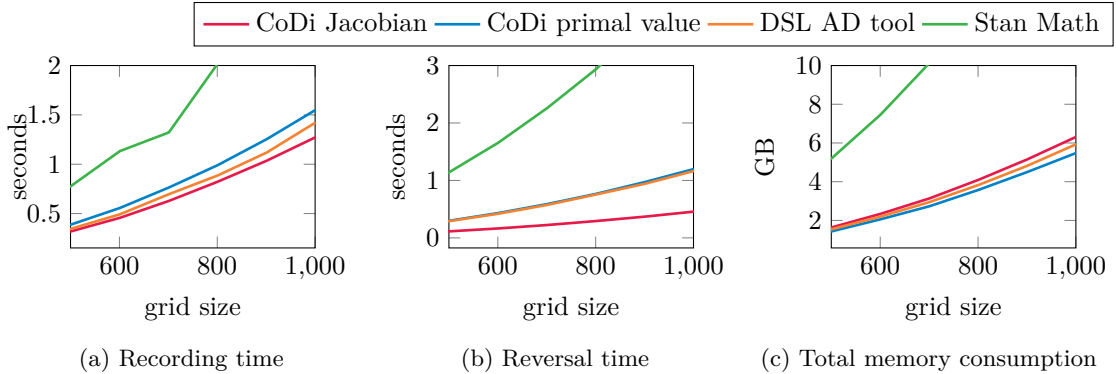
\begin{figure*}
  \begin{subfigure}{0.3\textwidth}
    \center
    \begin{tikzpicture}
      \begin{axis}[
        height=4cm,
        width=\textwidth,
        xlabel={grid size},
        xmin=500,
        xmax=1000,
        ymax=2,
        ylabel={seconds},
        legend columns=4,
        y label style={yshift=-1.5em},
        legend style={at={(0.5,1.1)},anchor=south west, cells={line width=1pt}},
      ]
  
      \timelinePlot{0}{5}{CoDi Jacobian}{data/burgers/codi2_jacobianIndexMulti.dat}
      \timelinePlot{0}{5}{CoDi primal value}{data/burgers/codi2_primalIndexMulti.dat}
      \timelinePlot{0}{5}{DSL AD tool}{data/burgers/adTool_primalIndexMulti.dat}
      \timelinePlot{0}{5}{Stan Math}{data/burgers/stanMathLib_stanMathLib.dat}
      \end{axis}
    \end{tikzpicture}
    \caption{Recording time}
  \end{subfigure}
  \begin{subfigure}{0.3\textwidth}
      \center
      \begin{tikzpicture}
        \begin{axis}[
          height=4cm,
          width=1\textwidth,
          xlabel={grid size},
          xmin=500,
          xmax=1000,
          ymax=3,
          ylabel={seconds},
          y label style={yshift=-1.5em},
          legend style={at={(1.,1.)},anchor=south east},
        ]
    
        \timelinePlot{0}{8}{CoDiPack2 jacobianIndexMulti}{data/burgers/codi2_jacobianIndexMulti.dat}
        \timelinePlot{0}{8}{CoDiPack2 primalIndexMulti}{data/burgers/codi2_primalIndexMulti.dat}
        \timelinePlot{0}{8}{adTool primalIndexMulti}{data/burgers/adTool_primalIndexMulti.dat}
        \timelinePlot{0}{8}{stanMathLib stanMathLib}{data/burgers/stanMathLib_stanMathLib.dat}
        \legend{}
        \end{axis}
      \end{tikzpicture}
      \caption{Reversal time}
    \end{subfigure}
    \begin{subfigure}{0.3\textwidth}
        \center
        \begin{tikzpicture}
          \begin{axis}[
            height=4cm,
            width=\textwidth,
            xlabel={grid size},
            xmin=500,
            xmax=1000,
            ymax=10,
            ylabel={GB},
            y filter/.code={\pgfmathparse{#1/1024.0}\pgfmathresult},
            legend columns=4, 
            y label style={yshift=-1em},
            legend style={at={(1.,1.)},anchor=south east},
          ]
      
          \timelinePlot{0}{18}{CoDi Jacobian}{data/burgers/codi2_jacobianIndexMulti.dat}
          \timelinePlot{0}{18}{CoDi primal value}{data/burgers/codi2_primalIndexMulti.dat}
          \timelinePlot{0}{18}{DSL AD tool}{data/burgers/adTool_primalIndexMulti.dat}
          \timelinePlot{0}{18}{Stan Math}{data/burgers/stanMathLib_stanMathLib.dat}
          \legend{}
          \end{axis}
        \end{tikzpicture}
        \caption{Total memory consumption}
      \end{subfigure}
  \caption{Time and memory measurements for the Burgers' test case.}
  \label{fig:burgers_result}
\end{figure*}

\subsection{Linear algebra tests}

Since the performance on pure double computations is comparable for the new DSL AD tool, we want to explore the benefits on linear algebra expressions. The Eigen library \cite{eigenweb} is used for these implementations. The first two cases are the same as in the Eigen-AD paper \cite{eigenAD2020} for comparability. They are a matrix matrix multiplication (T1) and a linear system solve with a Householder QR decomposition (T2). The other two test cases are taken from \cite{genSmalLinAlg2018} and the associated git repository. The third case is the Kalman filter (T3) that is used in control systems \cite{scharf98}. The fourth is the convex L1 analysis solver (T4) which is used for image denoising \cite{becker2011templates}. The codes for T3 and T4 are shown in Appendix \ref{sec:code_t3_t4} in Figures \ref{fig:t3_code} and \ref{fig:t4_code}, respectively. All tests are run on the same node as the Burgers' equation and only one process is run on the entire node. The same tools are compared with the addition of a specialized version of the Stan Math Library for the first two test cases. Here, specialized methods \ic{multiply} and \ic{mdivide_left} are used.

The results for T1, matrix matrix multiplication, are shown in Figure \ref{fig:results_matmul}. The strength of the new DSL AD tool is very pronounced in this case. In addition, the specialized implementation of the Stan Math Library shows the importance of exploiting the structure. For both tools, only one operation is recorded on the tape and, in the case of the DSL AD tool, only the memory for the overwritten result is required. The enhancement in memory and time efficiency of the Stan Math Library, without the special operation, in comparison to CoDiPack, is due to an already specialized implementation for multiplications in Eigen. If CoDiPack would implement the Eigen-AD interface, similar improvements would be expected. These results confirm the findings in \cite{eigenAD2020}.

The results in Figure \ref{fig:results_solve} for the linear system solve (T2) show the same general behavior as the T1 case. Here, a difference in memory and time can be seen between the DSL AD tool and the specialized operation in the Stan Math Library. The time discrepancy in the reverse evaluation is due to an additional QR decomposition, and solve of the primal linear system. The Stan Math implementation retains the primal result and QR decomposition. It can then reuse these values. A future optimization in the DSL AD tool could also store these results, which would increase the performance here. The memory difference will need to be further investigated on larger applications. Since only one operation is recorded here, a closer analysis is not really possible.

\begin{figure*}[p]
  \begin{center}
  \begin{subfigure}{0.3\textwidth}
    \center
    \begin{tikzpicture}
      \begin{axis}[
        height=4cm,
        width=\textwidth,
        xlabel={Matrix size},
        xmin=400,
        xmax=1000,
        ymax=15,
        ylabel={seconds},
        legend columns=5,
        y label style={yshift=-1.5em},
        legend style={at={(-0.25,1.1)},anchor=south west, cells={line width=1pt}},
      ]
  
      \timelinePlot{0}{1}{CoDi Jacobian}{data/mat_mul/CoDiPack2_jacobianIndex.dat}
      \timelinePlot{0}{1}{CoDi primal value}{data/mat_mul/CoDiPack2_primalValueIndex.dat}
      \timelinePlotWithMark{0}{1}{DSL AD tool}{data/mat_mul/adTool_primalValueIndex.dat}
      \timelinePlot{0}{1}{Stan Math}{data/mat_mul/stanMathLib_val.dat}
      \timelinePlot{0}{1}{Stan Math (special op)}{data/mat_mul/stanMathLib_valmat.dat}
      \end{axis}
    \end{tikzpicture}
    \caption{Recording time}
  \end{subfigure}
  \begin{subfigure}{0.3\textwidth}
      \center
      \begin{tikzpicture}
        \begin{axis}[
          height=4cm,
          width=1\textwidth,
          xlabel={Matrix size},
          xmin=400,
          xmax=1000,
          ymax=10,
          ylabel={seconds},
          y label style={yshift=-1.5em},
          legend style={at={(1.,1.)},anchor=south east},
        ]
    
      \timelinePlot{0}{4}{CoDi Jacobian}{data/mat_mul/CoDiPack2_jacobianIndex.dat}
      \timelinePlot{0}{4}{CoDi primal value}{data/mat_mul/CoDiPack2_primalValueIndex.dat}
      \timelinePlotWithMark{0}{4}{DSL AD tool}{data/mat_mul/adTool_primalValueIndex.dat}
      \timelinePlot{0}{4}{Stan Math}{data/mat_mul/stanMathLib_val.dat}
      \timelinePlot{0}{4}{Stan Math (special op)}{data/mat_mul/stanMathLib_valmat.dat}
        \legend{}
        \end{axis}
      \end{tikzpicture}
      \caption{Reversal time}
    \end{subfigure}
    \begin{subfigure}{0.3\textwidth}
        \center
        \begin{tikzpicture}
          \begin{axis}[
            height=4cm,
            width=\textwidth,
            xlabel={Matrix size},
            xmin=400,
            xmax=1000,
            ymax=20,
            y filter/.code={\pgfmathparse{#1/1024.0}\pgfmathresult},
            ylabel={GB},
            legend columns=4, 
            y label style={yshift=-1em},
            legend style={at={(1.,1.)},anchor=south east},
          ]
      
          \timelinePlot{0}{14}{CoDi Jacobian}{data/mat_mul/CoDiPack2_jacobianIndex.dat}
          \timelinePlot{0}{14}{CoDi primal value}{data/mat_mul/CoDiPack2_primalValueIndex.dat}
          \timelinePlotWithMark{0}{14}{DSL AD tool}{data/mat_mul/adTool_primalValueIndex.dat}
          \timelinePlot{0}{14}{Stan Math}{data/mat_mul/stanMathLib_val.dat}
          \timelinePlot{0}{14}{Stan Math (special op)}{data/mat_mul/stanMathLib_valmat.dat}
          \legend{}
          \end{axis}
        \end{tikzpicture}
        \caption{Total memory consumption}
      \end{subfigure}
  \end{center}
  \caption{Time and memory measurements for the matrix matrix multiplication (T1) test case. The values of the DSL AD tool for a matrix size of 700 are: recording 0.059\,s, reversal 0.135\,s, memory 0.169\,GB.}
  \label{fig:results_matmul}
\end{figure*}
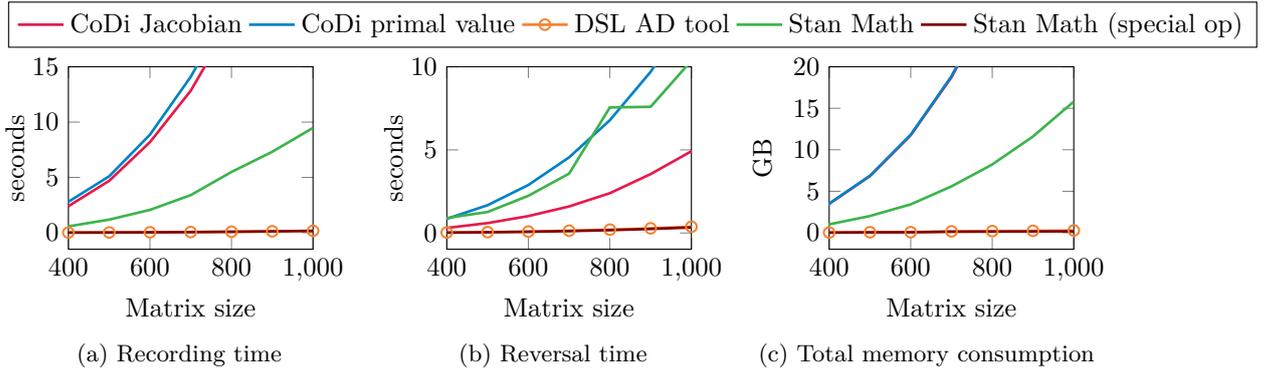
\begin{figure*}[p]
  \begin{center}
  \begin{subfigure}{0.3\textwidth}
    \center
    \begin{tikzpicture}
      \begin{axis}[
        height=4cm,
        width=\textwidth,
        xlabel={Matrix size},
        xmin=400,
        xmax=1000,
        ymax=5,
        ylabel={seconds},
        legend columns=5,
        y label style={yshift=-1.5em},
        legend style={at={(-0.25,1.1)},anchor=south west, cells={line width=1pt}},
      ]
  
      \timelinePlot{0}{1}{CoDi Jacobian}{data/solve/CoDiPack2_jacobianIndex.dat}
      \timelinePlot{0}{1}{CoDi primal value}{data/solve/CoDiPack2_primalValueIndex.dat}
      \timelinePlotWithMark{0}{1}{DSL AD tool}{data/solve/adTool_primalValueIndex.dat}
      \timelinePlot{0}{1}{Stan Math}{data/solve/stanMathLib_val.dat}
      \timelinePlot{0}{1}{Stan Math (special op)}{data/solve/stanMathLib_val_spec.dat}
      \end{axis}
    \end{tikzpicture}
    \caption{Recording time}
  \end{subfigure}
  \begin{subfigure}{0.3\textwidth}
      \center
      \begin{tikzpicture}
        \begin{axis}[
          height=4cm,
          width=1\textwidth,
          xlabel={Matrix size},
          xmin=400,
          xmax=1000,
          ymax=5,
          ylabel={seconds},
          y label style={yshift=-1.5em},
          legend style={at={(1.,1.)},anchor=south east},
        ]
    
      \timelinePlot{0}{4}{CoDi Jacobian}{data/solve/CoDiPack2_jacobianIndex.dat}
      \timelinePlot{0}{4}{CoDi primal value}{data/solve/CoDiPack2_primalValueIndex.dat}
      \timelinePlotWithMark{0}{4}{DSL AD tool}{data/solve/adTool_primalValueIndex.dat}
      \timelinePlot{0}{4}{Stan Math}{data/solve/stanMathLib_val.dat}
      \timelinePlot{0}{4}{Stan Math (special op)}{data/solve/stanMathLib_val_spec.dat}
        \legend{}
        \end{axis}
      \end{tikzpicture}
      \caption{Reversal time}
    \end{subfigure}
    \begin{subfigure}{0.3\textwidth}
        \center
        \begin{tikzpicture}
          \begin{axis}[
            height=4cm,
            width=\textwidth,
            xlabel={Matrix size},
            xmin=400,
            xmax=1000,
            ymax=10,
            y filter/.code={\pgfmathparse{#1/1024.0}\pgfmathresult},
            ylabel={GB},
            legend columns=4, 
            y label style={yshift=-1em},
            legend style={at={(1.,1.)},anchor=south east},
          ]
      
          \timelinePlot{0}{14}{CoDi Jacobian}{data/solve/CoDiPack2_jacobianIndex.dat}
          \timelinePlot{0}{14}{CoDi primal value}{data/solve/CoDiPack2_primalValueIndex.dat}
          \timelinePlotWithMark{0}{14}{DSL AD tool}{data/solve/adTool_primalValueIndex.dat}
          \timelinePlot{0}{14}{Stan Math}{data/solve/stanMathLib_val.dat}
          \timelinePlot{0}{14}{Stan Math (special op)}{data/solve/stanMathLib_val_spec.dat}
          \legend{}
          \end{axis}
        \end{tikzpicture}
        \caption{Total memory consumption}
      \end{subfigure}
  \end{center}
  \caption{Time and memory measurements for the linear system solve (T2) test case. The values of the DSL AD tool for a matrix size of 700 are: recording 0.055\,s, reversal 0.114\,s, memory 0.367\,GB.}
  \label{fig:results_solve}
\end{figure*}
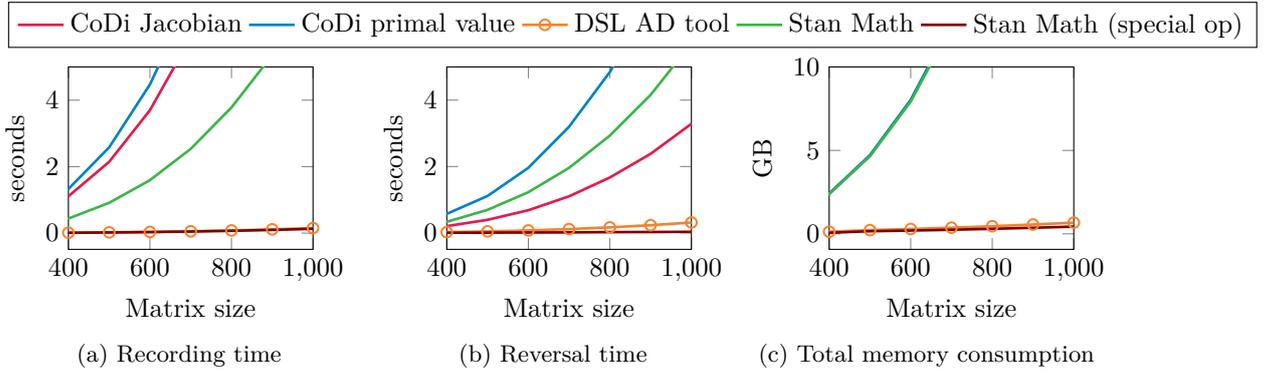
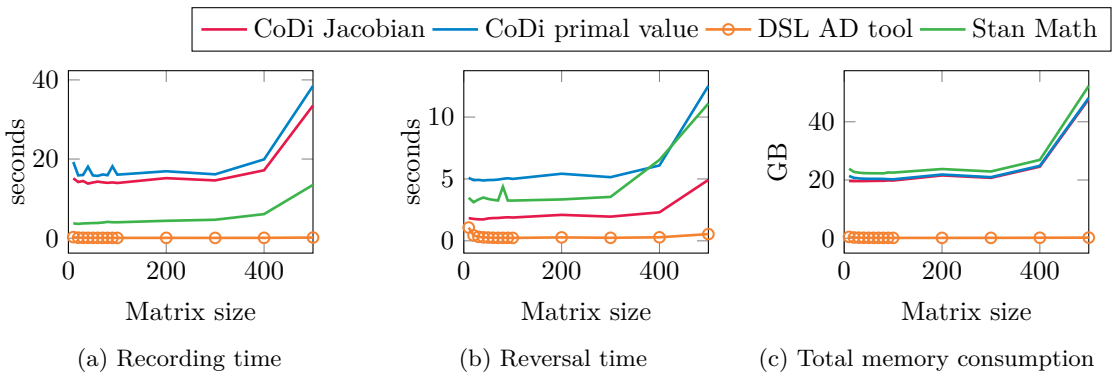
\begin{figure*}[p]
  \begin{center}
  \begin{subfigure}{0.3\textwidth}
    \center
    \begin{tikzpicture}
      \begin{axis}[
        height=4cm,
        width=\textwidth,
        xlabel={Matrix size},
        xmin=0,
        xmax=500,
        ylabel={seconds},
        legend columns=5,
        y label style={yshift=-1.5em},
        legend style={at={(0.5,1.1)},anchor=south west, cells={line width=1pt}},
      ]

      \timelinePlot{0}{1}{CoDi Jacobian}{data/KF/CoDiPack2_jacobianIndex.dat}
      \timelinePlot{0}{1}{CoDi primal value}{data/KF/CoDiPack2_primalValueIndex.dat}
      \timelinePlotWithMark{0}{1}{DSL AD tool}{data/KF/adTool_primalValueIndex.dat}
      \timelinePlot{0}{1}{Stan Math}{data/KF/stanMathLib_val.dat}
      \end{axis}
    \end{tikzpicture}
    \caption{Recording time}
  \end{subfigure}
  \begin{subfigure}{0.3\textwidth}
      \center
      \begin{tikzpicture}
        \begin{axis}[
          height=4cm,
          width=1\textwidth,
          xlabel={Matrix size},
        xmin=0,
        xmax=500,
          ylabel={seconds},
          y label style={yshift=-1.5em},
          legend style={at={(1.,1.)},anchor=south east},
        ]

      \timelinePlot{0}{4}{CoDi Jacobian}{data/KF/CoDiPack2_jacobianIndex.dat}
      \timelinePlot{0}{4}{CoDi primal value}{data/KF/CoDiPack2_primalValueIndex.dat}
      \timelinePlotWithMark{0}{4}{DSL AD tool}{data/KF/adTool_primalValueIndex.dat}
      \timelinePlot{0}{4}{Stan Math}{data/KF/stanMathLib_val.dat}
        \legend{}
        \end{axis}
      \end{tikzpicture}
      \caption{Reversal time}
    \end{subfigure}
    \begin{subfigure}{0.3\textwidth}
        \center
        \begin{tikzpicture}
          \begin{axis}[
            height=4cm,
            width=\textwidth,
            xlabel={Matrix size},
        xmin=0,
        xmax=500,
            y filter/.code={\pgfmathparse{#1/1024.0}\pgfmathresult},
            ylabel={GB},
            legend columns=4,
            y label style={yshift=-1em},
            legend style={at={(1.,1.)},anchor=south east},
          ]

          \timelinePlot{0}{14}{CoDi Jacobian}{data/KF/CoDiPack2_jacobianIndex.dat}
          \timelinePlot{0}{14}{CoDi primal value}{data/KF/CoDiPack2_primalValueIndex.dat}
          \timelinePlotWithMark{0}{14}{DSL AD tool}{data/KF/adTool_primalValueIndex.dat}
          \timelinePlot{0}{14}{Stan Math}{data/KF/stanMathLib_val.dat}
          \legend{}
          \end{axis}
        \end{tikzpicture}
        \caption{Total memory consumption}
      \end{subfigure}
  \end{center}
  \caption{Time and memory measurements for the Kalman Filter (T3) test case. The number of steps for each matrix size was chosen to create a 20\,GB for the CoDiPack Jacobian tape. The values of the DSL AD tool for a matrix size of 400 are: recording 0.077\,s, reversal 0.278\,s, memory 0.116\,GB.}
  \label{fig:results_KF}
\end{figure*}
\begin{figure*}
  \begin{center}
  \begin{subfigure}{0.3\textwidth}
    \center
    \begin{tikzpicture}
      \begin{axis}[
        height=4cm,
        width=\textwidth,
        xlabel={Matrix size},
        xmin=0,
        xmax=1000,
        ylabel={seconds},
        legend columns=5,
        y label style={yshift=-1.5em},
        legend style={at={(0.5,1.1)},anchor=south west, cells={line width=1pt}},
      ]

      \timelinePlot{0}{1}{CoDi Jacobian}{data/l1a/CoDiPack2_jacobianIndex.dat}
      \timelinePlot{0}{1}{CoDi primal value}{data/l1a/CoDiPack2_primalValueIndex.dat}
      \timelinePlotWithMark{0}{1}{DSL AD tool}{data/l1a/adTool_primalValueIndex.dat}
      \timelinePlot{0}{1}{Stan Math}{data/l1a/stanMathLib_val.dat}
      \end{axis}
    \end{tikzpicture}
    \caption{Recording time}
  \end{subfigure}
  \begin{subfigure}{0.3\textwidth}
      \center
      \begin{tikzpicture}
        \begin{axis}[
          height=4cm,
          width=1\textwidth,
          xlabel={Matrix size},
        xmin=0,
        xmax=1000,
          ylabel={seconds},
          y label style={yshift=-1.5em},
          legend style={at={(1.,1.)},anchor=south east},
        ]

      \timelinePlot{0}{4}{CoDi Jacobian}{data/l1a/CoDiPack2_jacobianIndex.dat}
      \timelinePlot{0}{4}{CoDi primal value}{data/l1a/CoDiPack2_primalValueIndex.dat}
      \timelinePlotWithMark{0}{4}{DSL AD tool}{data/l1a/adTool_primalValueIndex.dat}
      \timelinePlot{0}{4}{Stan Math}{data/l1a/stanMathLib_val.dat}
        \legend{}
        \end{axis}
      \end{tikzpicture}
      \caption{Reversal time}
    \end{subfigure}
    \begin{subfigure}{0.3\textwidth}
        \center
        \begin{tikzpicture}
          \begin{axis}[
            height=4cm,
            width=\textwidth,
            xlabel={Matrix size},
        xmin=0,
        xmax=1000,
            y filter/.code={\pgfmathparse{#1/1024.0}\pgfmathresult},
            ylabel={GB},
            legend columns=4,
            y label style={yshift=-1em},
            legend style={at={(1.,1.)},anchor=south east},
          ]

          \timelinePlot{0}{14}{CoDi Jacobian}{data/l1a/CoDiPack2_jacobianIndex.dat}
          \timelinePlot{0}{14}{CoDi primal value}{data/l1a/CoDiPack2_primalValueIndex.dat}
          \timelinePlotWithMark{0}{14}{DSL AD tool}{data/l1a/adTool_primalValueIndex.dat}
          \timelinePlot{0}{14}{Stan Math}{data/l1a/stanMathLib_val.dat}
          \legend{}
          \end{axis}
        \end{tikzpicture}
        \caption{Total memory consumption}
      \end{subfigure}
  \end{center}
  \caption{Time and memory measurements for the L1-analysis convex solver (T4) test case. The number of steps for each matrix size was chosen to create a 20\,GB for the CoDiPack Jacobian tape. The values of the DSL AD tool for a matrix size of 400 are: recording 0.061\,s, reversal 0.341\,s, memory 0.106\,GB.}
  \label{fig:results_L1a}
\end{figure*}
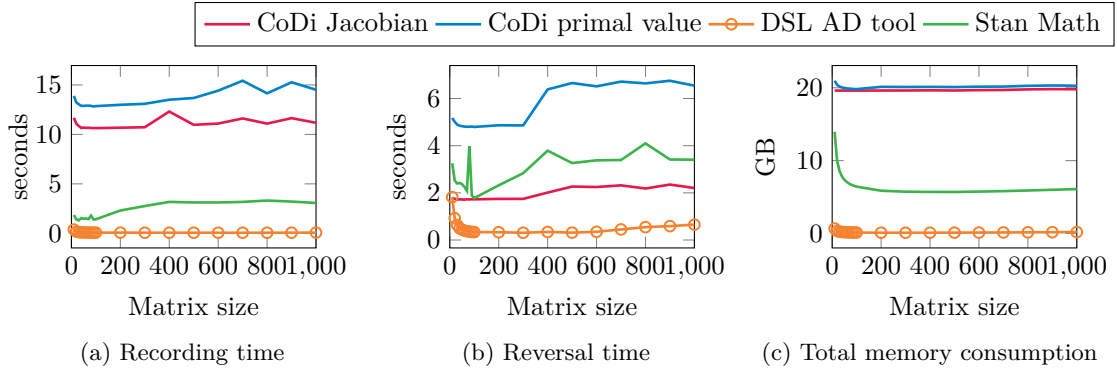

For T3 and T4, the presentation of the results is changed to a more realistic one. Multiple runs for the T3 and T4 kernels are recorded so that larger tape sizes are reached. This allows for a better comparison of the tools for small matrix sizes and stabilizes the results further. The number of steps for each dimension is chosen such that the CoDiPack Jacobian tape produces a tape of 20 GB. Then, all tools are run with the same number of steps for each dimension. The Kalman filter test (T3) consists of multiple linear systems solutions, matrix-matrix products, and matrix-vector products. Therefore, the test is similar to the T2 case, but shows the behavior of the tools if multiple operations are evaluated. The results in Figure \ref{fig:results_KF} show a mostly constant trend for the CoDiPack tools. The increase in the higher matrix dimension comes from the larger tape sizes. Here, one kernel evaluation creates a tape larger than 20 GB. The DSL AD tool can show its strength in this setting. The overhead for the recording and evaluation time is very low. Also, the overall memory consumption is quite low. For a matrix size of 10 the memory is 500 MB. Here, more old primal values need to be stored. For a matrix size of 400 the memory is 170 MB. In this case fewer kernel evaluations are performed and the DSL AD tool needs to store less old primal values. The benefits of the expression templates, where no intermediate results have to be stored, is more pronounced here.

The L1-analysis convex solver test (T4) uses mostly matrix-vector multiplication and vector operations. These operations benefit less from a specialization, which makes the test harder on the AD tools. The regular version of the Stan Math Library has already optimizations enabled for most of the operations in the test. Therefore, the memory difference with respect to CoDiPack is quite large in Figure \ref{fig:results_L1a}. A version of the Stan Math Library where only specialized routines are used leads to worse results. The DSL AD tool again shows very good performance. The factors with respect to a primal evaluation, which means that \icNC{double} is used as the underlying floating-point type, are also quite low. The factor for the recording has a maximum of 1.27 at a matrix size of 20. At a size of 200 it is down to 1.04 which is an overhead of 4\,\% for the recording. The reverse evaluation has a factor of 4.45 at a matrix size of 20. At a size of 200 it is down to 3.75, which is in the expected margins for the reverse mode of AD. The memory factor is 4.47 and 1.83 for a matrix size of 20 and 200 respectively. These are very good values and show that the underlying structure of the primal program is retained if the AD tool is specialized for the DSL the primal program uses.

One final observation can be made about runtime and memory complexity. The runtime complexity does not change for the test cases. The adjoint formulations have the same complexity as the primal algorithms. The difference is just the factor that is usually dropped in the $\mathcal{O}$-notation. The memory complexity goes down from $\mathcal{O}(n^3)$ to $\mathcal{O}(n^2)$ when switching from a pure operator overloading approach (CoDiPack) to a special handling (Stan Math). The DSL AD tool has the same memory complexity as the Stan Math library. Because of the different identifier handling approach and by storing fewer intermediate results (expression templates), the $\mathcal{O}$-factor is much lower.

\section{Conclusion and outlook}
\label{sec:conclusion}

In this paper, we present a new AD tool for DSLs. The tool uses a primal value taping approach and avoids the storing of intermediate values through expression templates. To this end, the expression templates are extended for lhs expression, for member operations, and for retaining temporaries during the reverse evaluation of one expression. The considerations behind the design of the DSL AD tool are described. These are the choices for a primal value taping approach, the data layout for the tape, the management of identifiers, and the handling of the primal and adjoint data. Annotations for the generation of expression templates for a DSL are highlighted. These have to be added to the source code of the DSL or can be provided in a separate header file. Afterwards, the expressions are generated from these annotations.

The performance tests in the coupled Burgers' equation show that the DSL AD tool has a performance similar to that of a CoDiPack primal value tape. Therefore, the generalization for DSLs does not need to affect the performance of the AD tool. The results of the synthetic tests T1 and T2 show that the DSL AD tool provides the same level of optimization as similar approaches in the Stan Math library or Eigen-AD. The tests T3 and T4 show a good performance of the DSL AD tool for smaller matrix dimensions. The runtime and memory factors presented for the T4 case show the benefits of an AD tool that retains the underlying structure of the primal program.

In order to show the true benefits of the approach, it needs to be applied to real world applications. Here, floating-point computations are usually interleaved with DSL computations. Depending on how many of the operations are written with a DSL the performance of the DSL AD tool will either be as in the synthetic benchmarks or more in the region of the Burgers' test case. Furthermore, the annotation of the Eigen library needs to be completed so that the full range of linear algebra functions is available. Annotations for complex numbers and SIMD vectors should also be created.

\section*{Acknowledgments}
The authors would like to thank all reviewers for their review and helpful comments.

\begin{appendices}
\section{\emph{Hello, world!} example for the DSL AD tool}
\label{sec:hello_world}
\begin{codeNum}
  #include <adTool.hpp>
  #include <adTool_double.hpp>
  #include <iostream>
  
  using Real = adTool::ActiveValue<double, int>;
  
  int main() {
    adTool::PrimalValueTape& tape = adTool::getPrimalValueTape();
    tape.registerValueKind<double>();
    tape.setActive();
  
    Real a = 4.0;
    tape.registerInput(a);
    Real w = a * a;
    tape.registerOutput(w);
  
    tape.setPassive();
  
    w.setGradient(1.0);
    tape.evaluate();
    std::cout << "dw/da = " << a.getGradient() << std::endl;
    
    return 0;
  }
\end{codeNum}

\section{Algorithms for T3 and T4 test case}
\label{sec:code_t3_t4}
\algrenewcommand\algorithmicrequire{\textbf{Input:}}
\algrenewcommand\algorithmicensure{\textbf{Output:}}
  \begin{figure}[H]
    \begin{algorithmic}[t]
      \Require $F, B, Q, H, R, P, u, x, z$
      \Ensure $P, x$
      \State $y = F * x + B * u$;
      \State $Y = F * P * F^T + Q$;
      \State $v0 = z - H * y$;
      \State $M_1 = H * Y$;
      \State $M_2 = Y * H^T$;
      \State $M_3 = M _1 * H^T + R$;
      \State $\underline{U}^T * \underline{U} = M_3$;
      \State $U^T * \underline{v_1} = v0$;
      \State $U * \underline{v_2} = v_1$;
      \State $U^T * \underline{M_4} = M_1$;
      \State $U * \underline{M_5} = M_4$;
      \State $x = y + M_2 * v_2$;
      \State $P = Y - M_2 * M_5$;
    \end{algorithmic}
    \caption{Kalman Filter (T3): Test cases and code taken from \cite{genSmalLinAlg2018}.}
    \label{fig:t3_code}
  \end{figure}
  \begin{figure}[H]
    \begin{algorithmic}
      \Require $W , A, x_0, y, v_1, z_1, v_2, z_2, \alpha, \beta, \tau$
      \Ensure $v_1, z_1, v_2, z_2$
      \State $y_1 = \alpha * v_1 + \tau * z_1$;
      \State $y_2 = \alpha * v_2 + \tau * z_2$;
      \State $x_1 = W^T * y_1 - A^T * y_2$;
      \State $x = x_0 + \beta * x_1$;
      \State $z_1 = y_1 - W * x$;
      \State $z_2 = y_2 - (y - A * x)$;
      \State $v_1 = \alpha * v_1 + \tau * z_1$;
      \State $v_2 = \alpha * v_2 + \tau * z_2$;
    \end{algorithmic}
    \caption{L1-analysis convex solver (T4): Test cases and code taken from \cite{genSmalLinAlg2018}.}
    \label{fig:t4_code}
  \end{figure}
\end{appendices}

\bibliographystyle{siam}
\bibliography{citations}

\end{document}